\documentclass[a4paper,11pt]{article}
\usepackage{jinstpub} % for details on the use of the package, please see the JINST-author-manual
\usepackage{lineno}
\usepackage{adjustbox}

\title{\boldmath Luminosity performance of SuperKEKB}

\author[a,b,1]{D. Zhou\note{Corresponding author.}}
\emailAdd{dmzhou@post.kek.jp}
\author[a]{K. Ohmi}
\author[a]{Y. Funakoshi}
\author[a,b]{and Y. Ohnishi}
\affiliation[a]{KEK, 1-1 Oho, Tsukuba 305-0801, Japan}
\affiliation[b]{The Graduate University for Advanced Studies, SOKENDAI}

\abstract{Since April 2020, the SuperKEKB has been operating with the crab waist scheme. The luminosity record achieved in June 2022 was $4.71 \times 10^{34} \text{ cm}^{-2}\text{s}^{-1}$, which overtook its predecessor KEKB by more than a factor of 2. The beam-beam interaction plays a key role in causing vertical blowup and consequently limiting the luminosity performance of SuperKEKB. In this paper, we examine luminosity tunings under the influence of beam-beam effects and review the luminosity performance of SuperKEKB with the crab-waist operation from 2020 to 2022.}


\begin{document}
\maketitle
\flushbottom

\section{Introduction}
\label{sec:intro}

%Ref.~\cite{furman1994beam}: First-order beam-beam effects: Stopbands, tune shifts, dynamic beta, and dynamic emittance. Beam-beam tune shifts and beam-beam parameter are different.

SuperKEKB~\cite{Ohnishi2013PTEP} is an asymmetric electron-positron collider under operation at KEK. The ``nano-beam'' scheme was chosen for the collision with a large crossing angle of 83 mrad and small vertical beta functions $\beta_y^*$ =0.27/0.3 mm for the positron/electron beams. The crab waist~\cite{raimondi2007beam} was optional for the baseline design configuration~\cite{SuperKEKBTDR} of SuperKEKB. Beam commissioning until the early stage of Phase-3 showed that the vertical blowup driven by beam-beam interaction was the major show-stopper in achieving high luminosity. As a countermeasure, the compact crab waist scheme~\cite{oide2016design} of FCC-ee was introduced to SuperKEKB in April 2020. Since then, SuperKEKB has been breaking the luminosity record of e+e- colliders. The latest luminosity record was $4.71 \times 10^{34} \text{ cm}^{-2}\text{s}^{-1}$, achieved at SuperKEKB on Jun. 22, 2022. In this paper, we review the luminosity performance of SuperKEKB from 2020 to 2022 by examining the luminosity tunings. For further details of beam-beam issues, the reader is referred to Ref.~\cite{zhou2023bb} and references therein.

The paper is organized as follows. The beam-beam parameter and its relation with luminosity are clarified in different cases in Sec.~\ref{sec:notation}. The process of luminosity tuning at SuperKEKB is presented in detail in Sec.~\ref{sec:lumtuning}. In Sec.~\ref{sec:lumperformance}, the achieved luminosity performance in different phases of SuperKEKB is presented and compared with that of KEKB and of the SuperKEKB design goal. The discrepancy in luminosity between the simulation and the real machine is briefly introduced in Sec.~\ref{sec:lsppuzzle} for both SuperKEKB and KEKB. In the end, we give a perspective of achieving the target luminosity at SuperKEKB from the beam-beam viewpoint in Sec.~\ref{sec:bbperspective} and summarize the paper in Sec.~\ref{sec:summary}.

\section{Notation}
\label{sec:notation}

The ``nano-beam'' scheme assumes collision with a large Piwinski angle
\begin{equation}
    \Phi_P=\frac{\sigma_z}{\sigma_x^*}\tan\frac{\theta_c}{2} \gg 1,
\end{equation}
and hourglass condition
\begin{equation}
    \frac{\beta_{y}^*} {\sigma_{x}^*} \tan\frac{\theta_c}{2} \gtrsim 1.
\end{equation}
Here we assume symmetric beams for simplification of the discussion. The quantities included are $\sigma_z$ the bunch length, $\theta_c$ the full crossing angle, $\sigma_x^*$ the horizontal beam size at the interaction point (IP), and $\beta_y^*$ the vertical beta function at the IP. With the above conditions satisfied, simple scaling laws are good enough to discuss luminosity and beam-beam parameters for the case of SuperKEKB. With hourglass effects negligible, the luminosity can be well approximated by
\begin{equation}
    \mathcal{L}\approx \frac{N_bN_+N_-f}{2\pi \Sigma_y \Sigma_z \tan \frac{\theta_c}{2}}
e^{-\frac{\Delta_y^2}{2\Sigma_y^2}},
\label{eq:lum1}
\end{equation}
where $\Sigma_y=\sqrt{\sigma_{y+}^{*2}+\sigma_{y-}^{*2}}$ and $\Sigma_z=\sqrt{\sigma_{z+}^{2}+\sigma_{z-}^{2}}$ are the effective beam sizes at the IP, $f$ is the revolution frequency, $N_b$ is the number of bunches for collision, $N_\pm$ the bunch population, $\sigma_y^*$ is the vertical beam size at the IP, and $\Delta_y$ is the relative vertical orbit offset of the colliding beams at the IP. The subscripts $+/-$ represent the e+/e- beams, respectively. The crab waist tilts the beams' density distribution around the IP, consequently affecting the luminosity. For the ``nano-beam'' scheme, the crab waist increases the luminosity by a few percent at SuperKEKB~\cite{zhou2022formulae}. Since it is a small gain, we will also use Eq.~(\ref{eq:lum1}) to discuss the luminosity performance with crab-waist collision. The specific luminosity is defined as
\begin{equation}
    \mathcal{L}_{sp}
    =\frac{\mathcal{L}}{N_bI_{b+}I_{b-}}
    \approx \frac{1}{2\pi e^2f \Sigma_y \Sigma_z\tan \frac{\theta_c}{2}},
    \label{eq:lsp1}
\end{equation}
with $I_b$ the bunch current.

In the literature, there are alternative definitions of beam-beam parameters. The incoherent beam-beam parameter is defined as~\cite{hirata2013bb}
\begin{equation}
    \xi_{u\pm}^i =
    \frac{r_e}{2\pi\gamma_\pm}
    \frac{N_\mp\beta_{u\pm}^*}{\sigma_{u\mp}^*(\sigma_{u\mp}^*+\sigma_{y\mp}^*)},
    \label{eq:xi1}
\end{equation}
with the subscript $u$=$x$ or $y$ for the horizontal or vertical direction, respectively. This definition does not include the hourglass effect. For collision with a horizontal crossing angle, $\sigma_{x\mp}^*$ should be replaced by the effective horizontal beam size at the IP as
\begin{equation}
    \sigma_{x\mp}=\sqrt{\sigma_{z\mp}^2\tan^2\frac{\theta_c}{2}+\sigma_{x\mp}^{*2}}.
\end{equation}
For the ``nano-beam'' scheme with flat beams (i.e., $\sigma_y^*\ll \sigma_x^*$) for collision, there is $\xi_x\ll \xi_y$. In this case, the vertical beam-beam parameter is of importance for measuring the luminosity performance and can be well approximated by
\begin{equation}
    \xi_{y\pm}^i \approx \frac{r_e}{2\pi\gamma_\pm}
    \frac{N_\mp\beta_{y\pm}^*}{\sigma_{y\mp}^*\sqrt{\sigma_{z\mp}^2\tan^2\frac{\theta_c}{2}+\sigma_{x\mp}^{*2}}}.
    \label{eq:xi1_approx}
\end{equation}
Empirically, we can define the vertical beam-beam parameters of flat beams from luminosity as~\cite{ohmi2004luminosity}
\begin{equation}
    \mathcal{L} = \frac{1}{2er_e} \frac{\gamma_\pm I_\pm}{\beta_{y\pm}^*} \xi_{y\pm}^L,
    \label{eq:xi2_by_lum}
\end{equation}
with $I_\pm$ the total beam currents. Since the luminosity depends on the effective beam sizes of the two beams, this definition is relevant to the coherent beam-beam tune shifts~\cite{hirata2013bb}. Only when the hourglass effects are negligible and the colliding beams have symmetric beam sizes, there is $\xi_{y\pm}^i=\xi_{y\pm}^L$~\cite{zhou2022formulae}. This equality is useful in machine tunings for luminosity optimization as will be addressed later.

When the hourglass effects are taken into account, the incoherent beam-beam parameter can be written as
\begin{equation}
    \xi_{u\pm}^{ih}=\xi_{u\pm}^i R_{\xi u}^\pm,
    \label{eq:xi_hourglass}
\end{equation}
where the superscript $ih$ indicates the incoherent tune shift including hourglass effects, and $R_{\xi u}^\pm$ can be taken as the hourglass factor for the beam-beam tune shifts. $\xi_{u\pm}^{ih}$ can be calculated by integrating the beam-beam force felt by the on-axis particles when the beam distributions at the IP are known.

With the above formulations, the luminosity and the incoherent beam-beam parameters are correlated as
\begin{equation}
    \mathcal{L} = \frac{1}{2e r_e} \frac{\gamma_\pm I_\pm}{\beta_{y\pm}^*} 
      \frac{2\sigma_{y\pm}^*\sigma_{x\pm}}{\Sigma_y\Sigma_x} \xi_{y\pm}^{ih} \frac{R_\mathcal{L}}{R_{\xi y}^\pm},
      \label{eq:lum_by_xi1}
\end{equation}
where $\Sigma_x = \sqrt{\sigma_{x+}^{2}+\sigma_{x-}^{2}}$ is the horizontal effective beam size of the two beams with the crossing angle counted, and $R_\mathcal{L}$ is the hourglass reduction factor for luminosity. When the colliding beams have symmetric beam sizes, Eq.~(\ref{eq:lum_by_xi1}) reduces to
\begin{equation}
    \mathcal{L} = \frac{1}{2e r_e} \frac{\gamma_\pm I_\pm}{\beta_{y\pm}^*} 
    \xi_{y\pm}^{ih} \frac{R_\mathcal{L}}{R_{\xi y}^\pm}.
      \label{eq:lum_by_xi2}
\end{equation}
This formulation was routinely used at KEKB~\cite{kekbtdr}.

For the general formulations of luminosity and beam-beam tune shifts for flat-beam asymmetric colliders and their tests with the machine parameters of SuperKEKB, the reader is referred to Ref.~\cite{zhou2022formulae}.

\section{Luminosity tuning}
\label{sec:lumtuning}

The typical path of beam tuning to achieve a good luminosity delivered to the physics run at SuperkEKB is as follows
\begin{itemize}
    \item Systematic corrections of linear optics (closed orbit, beta functions, dispersions, and linear couplings) are done separately for the two rings after weekly maintenance with a total beam current of around 50 mA. Iterative optics corrections may be done to minimize the vertical single-beam emittance for both HER and LER.
    \item Using filling patterns for collision, an IP orbit feedback system is used to search the optimal orbit for the collision. After finding the collision (observed by the luminosity monitors of Belle II) orbit, manual scans of the closed orbit at IP are done to find the best position for collision, which is then locked by the IP orbit feedback system.
    \item RF phase (for fine-tuning of the beams' arrival time at IP) and the waists of vertical beta functions are scanned and optimized to ensure that the beams collide at their waist positions.
    \item The linear parameters at the IP (such as linear couplings, dispersions, etc.) are scanned (so-called IP knobs) in a prescribed loop to optimize the luminosity.
    \item After achieving a good luminosity, the major luminosity tuning is done, and the machine is switched to the physics run.
    \item During the physics run, iterative minor IP knobs (including chromatic couplings and chromatic dispersions) are frequently done by the shifters of the accelerator team to search for the best luminosity performance or to recover the luminosity from degradation. The luminosity degradation might be due to the slow drift of machine conditions (such as drifting of environment temperature, changes in hardware, etc.). For the minor IP knobs, the background of Belle II detectors and the measured luminosity are used as references for optimization.
\end{itemize}
\begin{figure}[htbp]
\centering
\includegraphics[width=.64\textwidth]{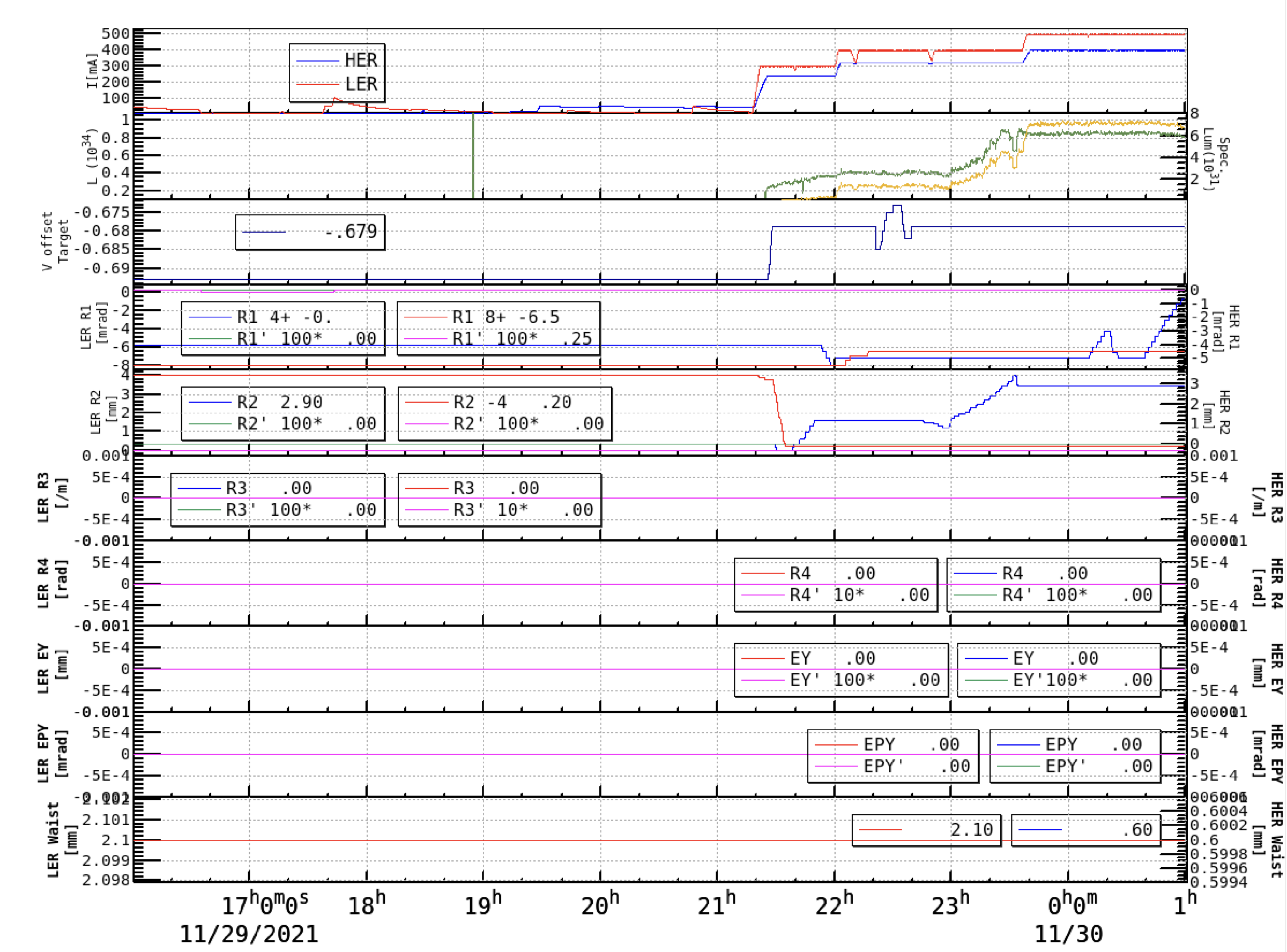}
\qquad
\includegraphics[width=.26\textwidth]{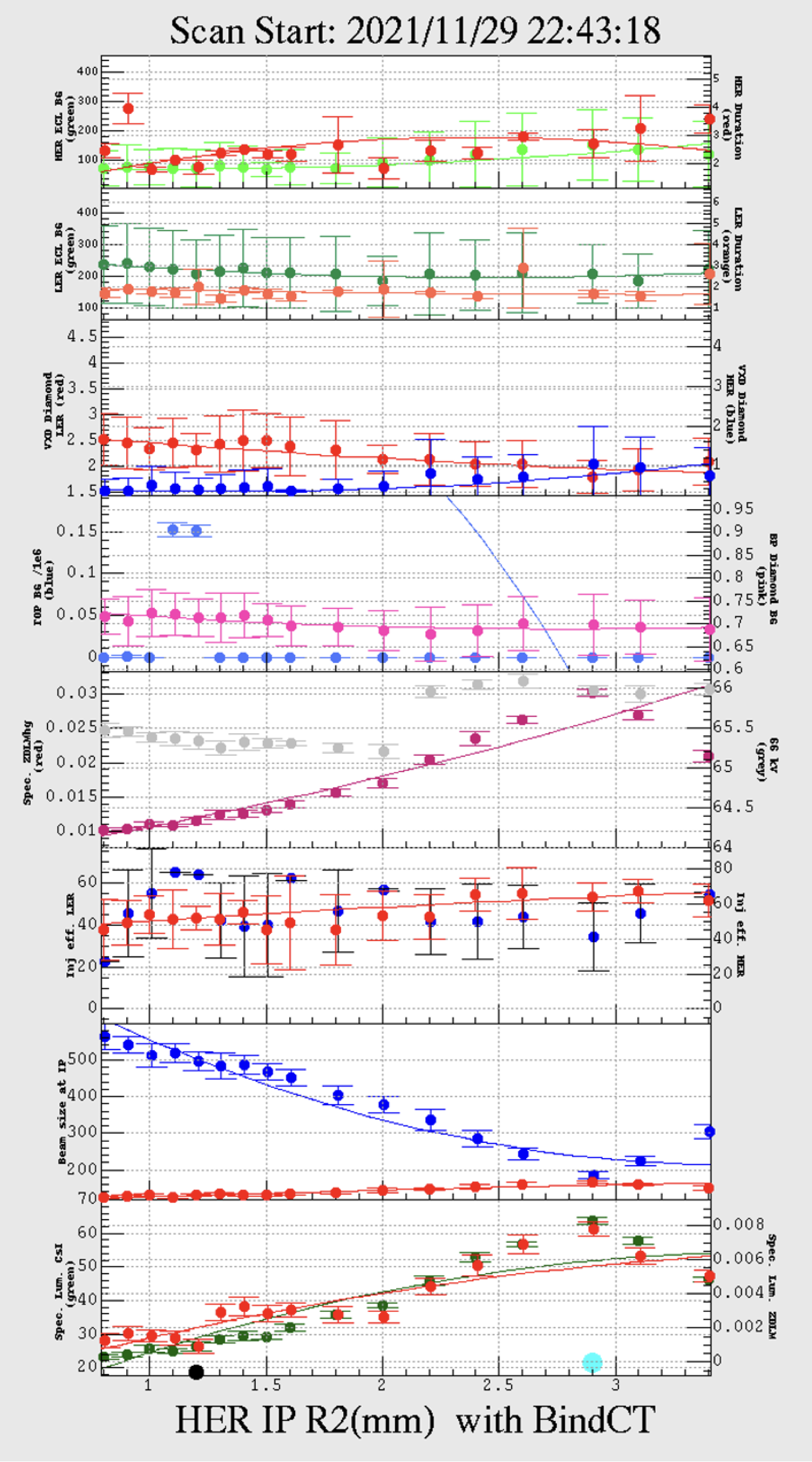}
\caption{Example of IP knobs for luminosity tuning at SuperKEKB.\label{fig:IPknobs}}
\end{figure}

Figure~\ref{fig:IPknobs} shows a typical example of luminosity tuning at SuperKEKB. The left figure presents the online data of machine parameters for 9 hours of beam tuning. From top to bottom, the subfigures indicate 1) the total currents of HER (blue) and LER (red) rings; 2) the luminosity (yellow) and specific luminosity (green); 3) the vertical relative orbit offset at the IP; 4) the linear coupling parameter $R_1$ at the IP (calculated from the online lattice models; for definitions of coupling parameters, see Ref.~\cite{SADwebpage}) of HER (blue) and LER (red), and the first-order chromaticity of $R_1$ of HER (green) and LER (magenta); 5) the linear coupling parameter $R_2$ and its chromaticity; 6) the linear coupling parameter $R_3$ and its chromaticity; 7) the linear coupling parameter $R_4$ and its chromaticity; 8) the vertical dispersion and its chromaticity at the IP; 9) the vertical dispersion prime and its chromaticity at the IP; 10) the waist position of HER (blue) and LER (red).

The timeline of Fig.~\ref{fig:IPknobs} is illustrated as follows. The LER optics corrections started at around 17:40 PM, and the HER optics corrections started at around 19:30 PM. The collision tuning started at around 21:20 PM by optimizing the vertical orbit offset $\Delta_y$ (Since the vertical beam sizes at the IP are in the order of 0.1 $\mu$m, the luminosity is very sensitive to $\Delta_y$ as shown in Eq.~(\ref{eq:lum1})). After finding the optimal beam orbit at the IP, the linear couplings at the IP $R_1$ and $R_2$ of the two rings were scanned one by one. Among the linear and chromatic parameters at the IP, the coupling parameters $R_1$ and $R_2$ are most sensitive to the vertical beam sizes at the IP. It should be emphasized that the linear optics corrections were mainly done globally. The couplings at the IP need further fine-tunings to ensure that they are corrected to zeros (Here, ``zeros'' means the luminosity is locally optimal but does not suggest the parameters are zeros from the online lattice model.). After these major IP knobs with moderate beam currents (It means that the major luminosity tuning is done with weak collective effects.), a good luminosity can usually be found, and the machine can be delivered to physics run. Aftward, the total beam currents will be increased gradually, and further luminosity optimization may be done under collective effects (such as beam-beam-driven beam-size blowups, impedance effects, etc.).

The right figure of Fig.~\ref{fig:IPknobs} shows a very effective IP knob of $R_2$ for HER (This knob started at 22:43 PM, seen in the fifth subfigure from the top in the left figure.). From top to bottom, the subfigures indicate 1) the background of the Belle II ECL detector~\cite{kovalenko2020jinst} from the HER beam (green) and the injection duration~\cite{herzberg2022thesis} of the HER beam (red); 2) the background of the Belle II ECL detector from the LER beam (green) and the injection duration of the LER beam (orange); 3) the background of the Belle II VXD diamond from the LER beam (red) and HER beam (blue); 4) the background of the Belle II TOP detector (blue) and BP diamond detector (pink); 5) the specific luminosity from ZDLM luminosity monitor (red)~\cite{pang2018ipac} and the 66 kV high-voltage power supply; 6) the injection efficiency of LER (red) and HER (blue); 7) the vertical beam size at the IP (X-ray monitors located far from the IP are used to measure the beam sizes. Using the optics functions at the X-ray monitors and at the IP of the designed lattice, the beam sizes at the IP are estimated via the relation of $\sigma_y=\sqrt{\beta_y\epsilon_y}$ with $\epsilon_y$ the vertical emittance.) of LER (red) and HER (blue); 8) the specific luminosity given by Belle II ECL/CsI detector (green) and ZDLM detector (red). It can be seen that the specific luminosity (see the fifth and eighth subfigures in the right figure of Fig.~\ref{fig:IPknobs}) is strongly correlated with the beam sizes (see the seventh subfigure in the right figure of Fig.~\ref{fig:IPknobs}), as expected from Eq.~(\ref{eq:lsp1}). With non-zero linear couplings at the IP, the beam-beam force causes extra vertical blowup, consequently leading to luminosity loss~\cite{zhou2010bb}. In principle, the best luminosity appears when the linear couplings at the IP are fully corrected to zeros. This can be extended to a general principle for luminosity tuning: The luminosity is a function of a parameter list $\Vec{R}=(R_1, R_2, ...)$. Scan the $i$-th parameter $R_i$(design) (We use the design lattice as a reference) and find the optimal condition
\begin{equation}
    \frac{\partial \mathcal{L}(\Vec{R})}{\partial R_i}=0,
\end{equation}
then we conclude the real value of $R_i$ is minimized though the design value of $R_i$ from the lattice is not. This criterion does not guarantee $R_i$(real) is fully corrected to zero since many other parameters might not be zeros (or, say, not optimized yet). Therefore, the shifters frequently scan the parameters of the list $\Vec{R}$ in the control room. Usually, the parameters, which are sensitive to luminosity (such as linear couplings $R_1$ and $R_2$ at the IP), are scanned with the highest priority. Nevertheless, luminosity tuning is a challenging task at SuperKEKB.

\section{Luminosity performance}
\label{sec:lumperformance}

From March 2018 to March 2020, SuperKEKB was operated without the crab waist. The beam-beam-driven resonances caused severe vertical beam-size blowups, limiting the luminosity performance. In April 2020, the crab waist scheme of Ref.~\cite{oide2016design} was implemented in SuperKEKB to suppress the beam-beam resonances. Since then, the luminosity performance has been improving. Table~\ref{tab:LuminosityComparison} (adapted from Table 1 of Ref.~\cite{funakohsi2022ipac}) compares the machine parameters of KEKB and SuperKEKB in four cases from left to right: 1) The machine parameters of KEKB leading to its best luminosity in June 2009; 2) The machine parameters of SuperKEKB in May 2020 when the crab waist started to be functional; 3) The machine parameters of SuperKEKB in June 2022 when the luminosity record of $4.71 \times 10^{34} \text{ cm}^{-2}\text{s}^{-1}$ was achieved; 4) The machine parameters of SuperKEKB design~\cite{SuperKEKBTDR}.
\begin{table}[htbp]
    \small
	\centering
	\caption{Comparison of KEKB and SuperKEKB machine parameters.}
	\label{tab:LuminosityComparison}
 \begin{adjustbox}{width=\textwidth}
 \begin{tabular}{p{0.2\textwidth} c c c c c c c c}
  \hline
  & \multicolumn{2}{c}{KEKB} & \multicolumn{2}{c}{SuperKEKB} &  \multicolumn{2}{c}{SuperKEKB} & \multicolumn{2}{c}{SuperKEKB} \\
  & \multicolumn{2}{c}{Achieved} & \multicolumn{2}{c}{2020 May 1st} & \multicolumn{2}{c}{2022 June 22nd} & \multicolumn{2}{c}{Design} \\
  \hline
  & LER & HER & LER & HER & LER & HER & LER & HER \\ \hline \hline
  $\rm I_{beam} [A]$ &1.637 &1.188 &0.438&0.517 &1.363 &
1.118 &3.6 &2.6 \\
 \# of bunches & \multicolumn{2}{c}{1585} & \multicolumn{2}{c}{783} &  \multicolumn{2}{c}{2249} & \multicolumn{2}{c}{2500} \\
$\rm I_{bunch} [mA]$ &1.033 &0.7495 &0.5593 & 0.6603 &0.606 &0.497 &1.440 &1.040 \\	
$\beta_y^*$ [mm] &5.9 &5.9 &1.0 & 1.0 &1.0 &1.0 &0.27 &0.30 \\
$\xi_y$ &0.129$^{a)}$ &0.090$^{a)}$ &0.0236$^{b)}$ &0.0219$^{b)}$ &0.0398$^{b)}$ & 0.0278$^{b)}$ &0.0881$^{c)}$ &0.0807$^{c)}$ \\
 & 0.10$^{b)}$ & 0.060$^{b)}$ & & &$0.0565^{d)}$ & $0.0434^{d)}$ & 0.069$^{b)}$ & 0.061$^{b)}$ \\
$\mathcal{L}$ [$\rm 10^{34}cm^{-2}s^{-1}$]& \multicolumn{2}{c}{2.11} & \multicolumn{2}{c}{1.57} & \multicolumn{2}{c}{4.71} & \multicolumn{2}{c}{80} \\
$\int \mathcal{L}dt$ [$\rm ab^{-1}$] & \multicolumn{2}{c}{1.04} & \multicolumn{2}{c}{0.03} & \multicolumn{2}{c}{0.424} & \multicolumn{2}{c}{50}
 \\\hline\hline
 \end{tabular}
 \end{adjustbox}
 \footnotesize{$^{a)}$Values of $\xi_y^{ih}$ calculated by Eq.~(\ref{eq:lum_by_xi2}); $^{b)}$Values of $\xi_y^L$ calculated by Eq.~(\ref{eq:xi2_by_lum}); $^{c)}$Values of $\xi_y^{ih}$ calculated by Eq.~(\ref{eq:xi_hourglass}); $^{d)}$Values in high bunch current study with 393 bunches by Eq.~(\ref{eq:xi2_by_lum}).}
\end{table}

The reader may notice that different formulae were used in Table~\ref{tab:LuminosityComparison} to calculate the beam-beam parameters at KEKB and SuperKEKB. For the KEKB and SuperKEKB design cases, the hourglass factor $R_\mathcal{L}/R_{\xi y}$ is around 0.7-0.8. Consequently, it is suitable to use $\xi_y^{ih}$ for a better estimate of beam-beam tune shift. For the current SuperKEKB with $\beta_y^*=1$ mm, the hourglass factor $R_\mathcal{L}/R_{\xi y}$ is very close to 1~\cite{zhou2022formulae}. In this case, it is convenient to use $\xi_y^L$, especially when the beam sizes at the IP cannot be accurately monitored. But, it is also noteworthy that $\xi_y^L$ will not be a good measure of the beam-beam tune shift felt by a beam when the beam sizes of the two beams are far from symmetric. During the operation of SuperKEKB, both $\xi_y^i$ (calculated from measured beam sizes using Eq.~(\ref{eq:xi1_approx})) and $\xi_y^L$ (calculated from measured luminosity using Eq.~(\ref{eq:xi2_by_lum})) are provided online to guide the luminosity tunings and optimizations. In other words, the ratio $\xi_y^L/\xi_y^i$ is used as an index for monitoring the symmetry of the colliding beams' sizes.

Further clarifications on Table~\ref{tab:LuminosityComparison} are given as follows.
\begin{itemize}
    \item The beam-beam parameters of HER and LER beams are not equal. This is because the luminosity is optimized without keeping the energy transparency condition $\gamma_+I_+=\gamma_-I_-$.
    \item The beam-beam parameters for the peak luminosity of $4.71 \times 10^{34} \text{ cm}^{-2}\text{s}^{-1}$ were lower than those achieved in high bunch machine study (In the machine study, less number of bunches but higher bunch currents were chosen.). This is because, during the physics run, there were high risks of machine failure (so-called sudden beam losses~\cite{ikeda2022eefact} and consequent damages to hardware in the rings) when the bunch currents exceeded about 0.7 mA in the LER. However, during the high bunch current machine studies, there was no limit in the bunch currents since the total currents were not as high in the physics run.
    \item The beam-beam parameters achieved in the high bunch current machine study are remarkably lower than the design values. This is due to the severe vertical beam-size blowup caused by beam-beam interaction and its interplay with other factors (such as impedance effects, nonlinear IR optics, imperfect crab waist, etc.). Investigations are ongoing to evaluate these effects quantitatively.
\end{itemize}

\begin{figure}[htbp]
\centering
\includegraphics[width=.8\textwidth]{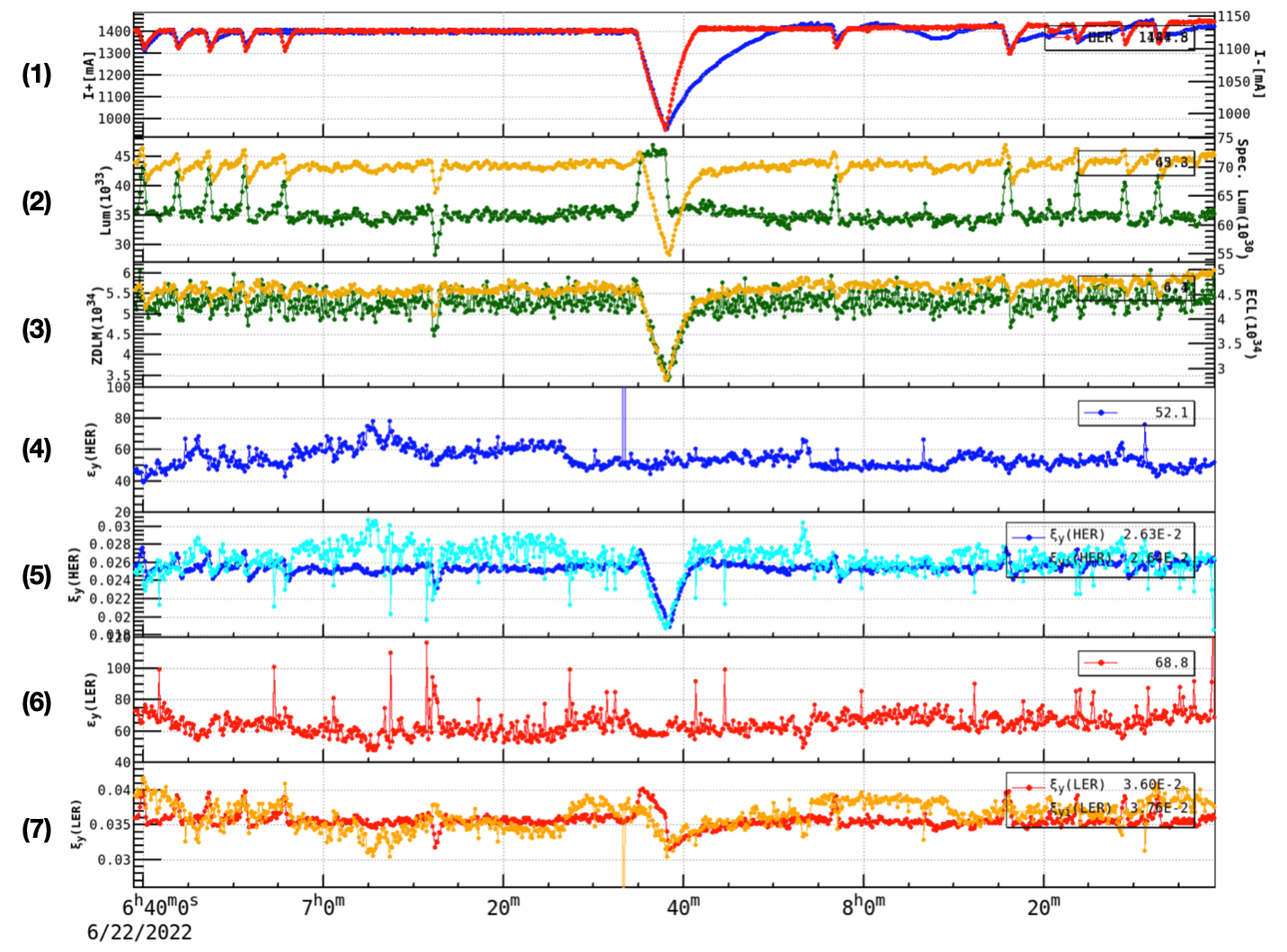}
\caption{Historical data in two hours for beam currents, luminosity, vertical emittance, and beam-beam parameters at SuperKEKB on Jun. 22, 2022. \label{fig:BB-Panel}}
\end{figure}

Figure~\ref{fig:BB-Panel} shows the machine parameters in 2 hours when the luminosity record of $4.71 \times 10^{34} \text{ cm}^{-2}\text{s}^{-1}$ was achieved. From top to bottom, the subfigures are illustrated as follows.
\begin{enumerate}
    \item[(1)] The total beam current of LER (red) and HER (blue). It is seen that the beam injection was intentionally stopped many times. The purpose is to achieve a peak luminosity, as seen in subfigure (2).
    \item[(2)] The total luminosity by ECL monitor (yellow, averaged with a 20-second window) and specific luminosity (green) calculated from the average ECL luminosity. A peak luminosity always appeared soon after the beam injection was stopped. This was because the ECL luminosity was affected by the background from beam injection~\cite{zhou2023bb}. The luminosity loss came from two sources: the LER injection background on ECL and the beam oscillation in LER excited by the injection kickers due to the imbalance of the upstream and downstream kickers' fields. This luminosity loss is more clearly seen in the specific luminosity. About 10\% of gain in specific luminosity can be seen when the injection is off. However, because of the short beam lifetime, only a small gain is seen in the total luminosity.
    \item[(3)] The total luminosity by ZDLM monitor (yellow) and the instant total luminosity by ECL monitor (green). The ZDLM luminosity monitor is less sensitive to beam injection background than the ECL monitor~\cite{zhou2023bb}.
    \item[(4)] The vertical emittance of HER beam measured by X-ray monitor. The vertical emittance is calculated from $\sigma_y=\sqrt{\beta_y\epsilon_y}$ with $\sigma_y$ measured by the X-ray monitor and $\beta_y$ given by design lattice.
    \item[(5)] The vertical beam-beam parameter calculated from ECL luminosity (blue), and the vertical beam-beam tune shift calculated from beam sizes (cyan) using Eq.~(\ref{eq:xi1}) for the electron beam (HER). As mentioned previously, the asymmetric beam sizes are the main sources of the discrepancy in the beam-beam parameters calculated from luminosity and measured beam sizes. With beam tunings, symmetric beam sizes may be achieved, and this discrepancy will be minimum. But the machine conditions constantly drift, making it difficult to keep the symmetric beam sizes for a long time.
    \item[(6)] The vertical emittance of LER beam measured by X-ray monitor.
    \item[(7)] The vertical beam-beam parameter calculated from ECL luminosity (red), and vertical beam-beam tune shift calculated from beam sizes (orange) for the positron beam (LER).
\end{enumerate}

\begin{figure}[htbp]
\centering
\includegraphics[width=.8\textwidth]{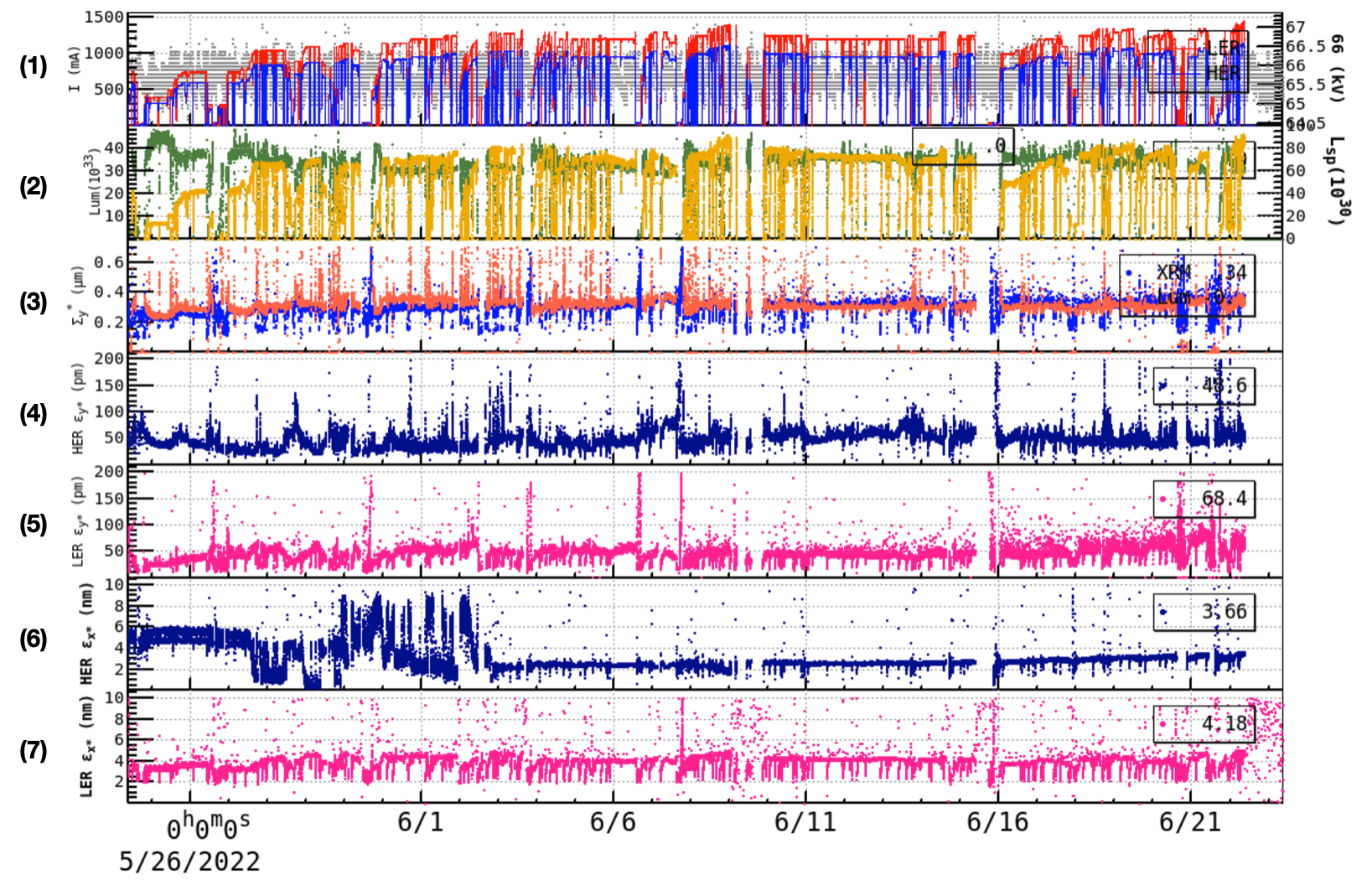}
\caption{Historical data in 30 days until Jun. 22, 2022, for beam currents, luminosity, vertical effective beam size $\Sigma_y$, and beam emittances at SuperKEKB. \label{fig:Lum-Panel-30days}}
\end{figure}

Figure~\ref{fig:Lum-Panel-30days} shows the machine parameters in 30 days until June 22, 2022 (the last day of the 2022b run of SuperKEKB). From top to bottom, the subfigures are described as follows.
\begin{enumerate}
    \item[(1)] The total beam current of LER (red) and HER (blue) and the 66 kV high-voltage power supply (gray).
    \item[(2)] The total luminosity (yellow) and specific luminosity (green) from the ECL luminosity monitor.
    \item[(3)] The vertical effective beam size $\Sigma_y$ at the IP, calculated from luminosity (blue) using Eq.~(\ref{eq:lsp1}) (the effective bunch length $\Sigma_z$ is a constant calculated by using the nominal bunch length of design lattices.) and from measured beam sizes (orange) using X-ray monitors. Overall, these two methods give similar results. The difference between the two values changed over time, depending on the beam tunings. Note that the bunch-current-dependent bunch lengthening was not considered in the luminosity method.
    \item[(4)] The vertical emittance of HER beam calculated from measurement data of X-ray monitor. The minimum value was around 20 pm, showing the best single-beam emittance (without collision) achieved at HER. With collision, beam-beam effects caused vertical blowup by a factor of more than 2 at the end of the 2022b run. Severe vertical blowups can be seen frequently, usually corresponding to non-optimal IP knobs or poor machine conditions for certain reasons.
    \item[(5)] The vertical emittance of LER beam calculated from measurement data of X-ray monitor. The minimum value was around 20 pm, showing the best single-beam emittance (without collision) achieved at LER. Similar to the HER beam, the beam-beam effects caused vertical blowup by a factor of more than 2. When pushing the total beam current to achieve higher luminosity, the vertical blowup in LER became more severe than in HER. This showed a hint of an interplay between beam-beam effects and vertical impedance effects in the LER. In practice, on June 21, 2022, the vertical collimator settings and vertical tune of LER were optimized to increase the threshold of single-beam vertical blowup (from around 0.5 mA to around 0.9 mA in bunch current). This also contributed to achieving the new luminosity record on Jun 22, 2022.
    \item[(6)] The horizontal emittance of HER beam calculated from measurement data of synchrotron radiation (SR) monitor. The data before June 3, 2022, were not reliable because the mirror of the SR monitor was broken. Weak horizontal blowup seemed to be seen from the measurement data, which is also predicted by beam-beam simulations~\cite{zhou2023bb}.
    \item[(7)] The horizontal emittance of the LER beam calculated from measurement data of the SR monitor. The measurement data showed clear horizontal blowups, which were sensitive to the horizontal tune of LER. This horizontal blowup is due to beam-beam effects and their interplay with beam-coupling impedance and was identified by simulations~\cite{zhou2023bb}.
\end{enumerate}

\begin{figure}[htbp]
\centering
\includegraphics[width=.8\textwidth]{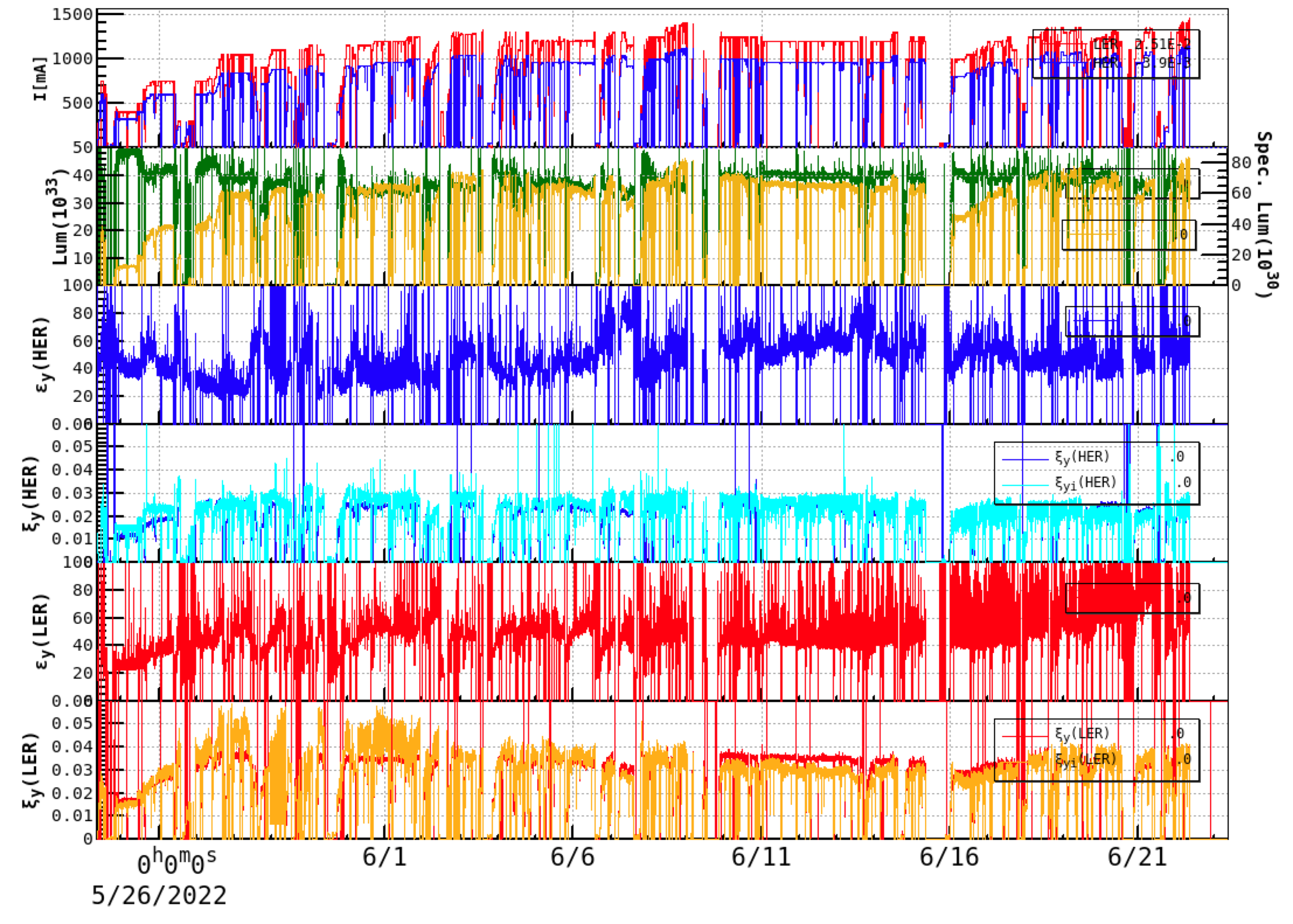}
\caption{Historical data in 30 days until Jun. 22, 2022, for beam currents, luminosity, vertical emittances, and beam-beam parameters at SuperKEKB. \label{fig:BB-Panel-30days}}
\end{figure}

Figure~\ref{fig:BB-Panel-30days} shows historical data in the same period as Fig.~\ref{fig:Lum-Panel-30days}. During the physics run, the achieved beam-beam parameters are around 0.04 and 0.03 for the LER and HER beams, respectively. When machine tunings were well done, the beam-beam parameters estimated from luminosity and XRM data had the best agreements.

\section{The puzzle of specific luminosity slope}
\label{sec:lsppuzzle}

\begin{figure}[htbp]
\centering
\includegraphics[width=.5\textwidth]{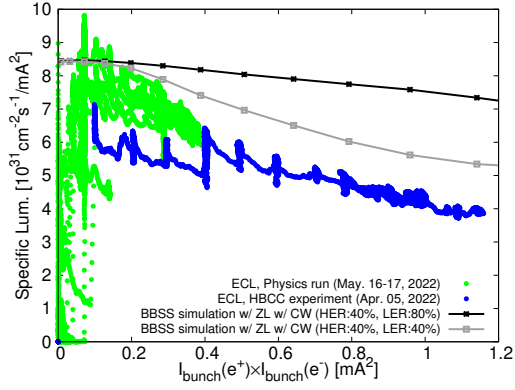}
\caption{Comparison of specific luminosity by predictions of simulations using BBSS code and by measurements of ECL luminosity monitor at SuperKEKB. \label{fig:lsp_comparison}}
\end{figure}

As shown in Eq.~(\ref{eq:lsp1}), the specific luminosity is a geometric parameter that defines the potential of extracting luminosity from a collider. For SuperKEKB, intensive investigations have been done to understand the luminosity performance, but a large discrepancy exists between simulations and measurements. This can be seen in Fig.~\ref{fig:lsp_comparison}. The green dots give the specific luminosity from the ECL monitor during the physics run with high total currents. The blue dots are data extracted from the high-bunch current collision (HBCC). At $I_{b+}I_{b-}<0.4 \text{ mA}^2$, the specific luminosity is higher than that from the HBCC machine study. This is because beam tunings were not done during the HBCC machine study due to limited beam time. The beam-beam simulations were done using BBSS code~\cite{ohmi2000bbss} with beam coupling impedance included. The simulations show two cases: one is for 40\% and 80\% of full crab waist strengths for HER and LER, respectively (this is the current configuration in machine operation); another is 40\% for both HER and LER. The measured specific luminosity as a function of bunch current product drops much faster than simulations, and it remains a puzzle at SuperKEKB. For further discussions on the sources of luminosity degradation, the reader is referred to Ref.~\cite{zhou2023bb}.

\begin{figure}[htbp]
\centering
\includegraphics[width=.5\textwidth]{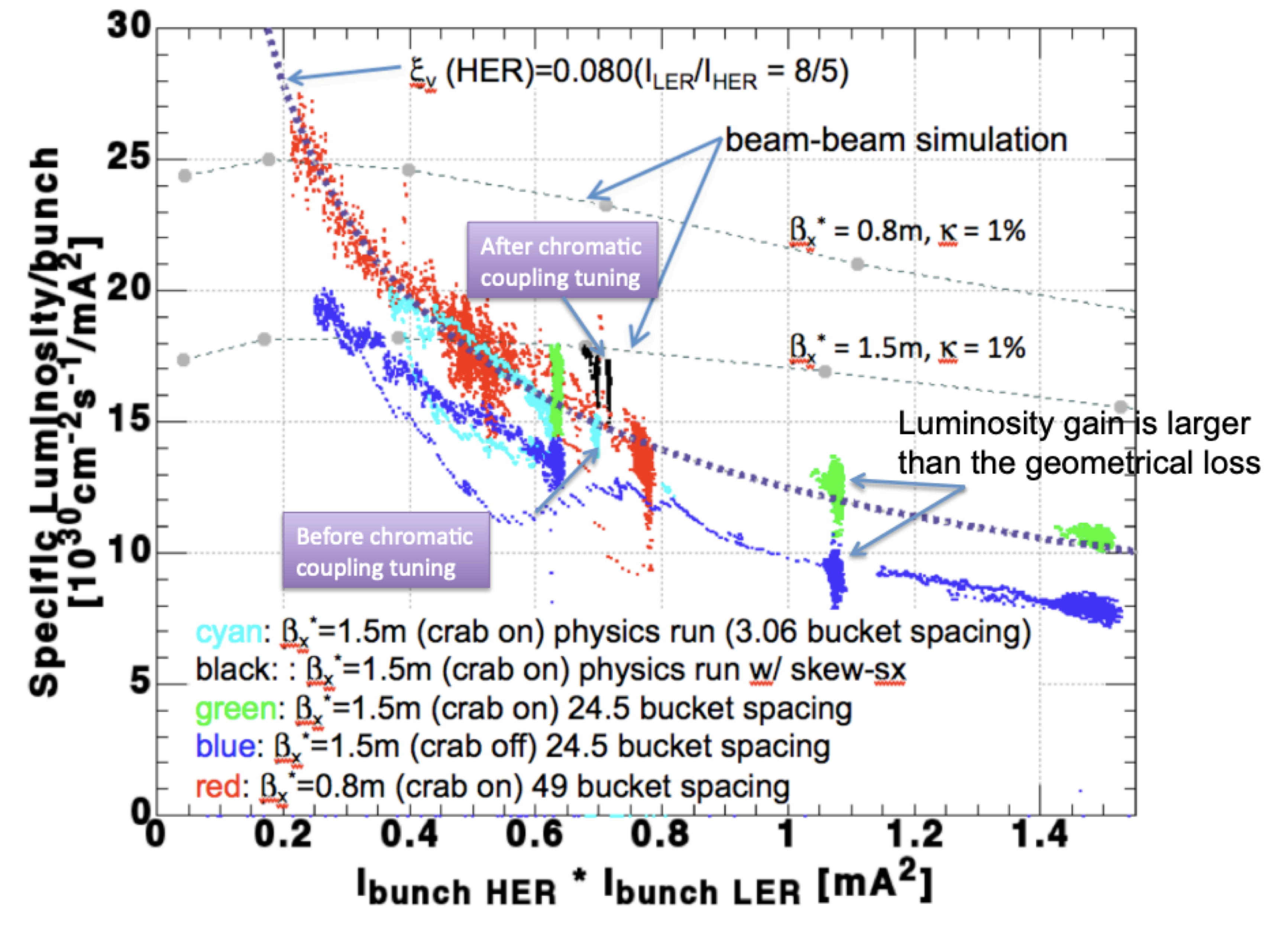}
\caption{Comparison of specific luminosity by predictions of simulations and by measurements at KEKB. \label{fig:lsp_comparison_kekb}}
\end{figure}

The steep slope of measured specific luminosity also existed in KEKB, which beam-beam simulations could not explain. This can be seen in Fig.~\ref{fig:lsp_comparison_kekb} (adapted from Ref.~\cite{funakoshi2010kekb}). The crab cavities were installed, and remarkable luminosity gain was achieved (see the difference between the green and blue dots) at KEKB. Skew sextupoles were installed to correct the chromatic couplings at the IP and contributed to extra luminosity gain (see the black dots). However, all these did not change the steep slope of measured specific luminosity.

%\appendix
%\section{Some title}
%Please always give a title also for appendices.

\section{Beam-beam perspective on achieving target luminosity at SuperKEKB}
\label{sec:bbperspective}

The SuperKEKB is targeting a peak luminosity of $1 \times 10^{35} \text{ cm}^{-2}\text{s}^{-1}$ after the long-term shutdown 1 (LS1, from July 2022 to the end of 2023) and $6 \times 10^{35} \text{ cm}^{-2}\text{s}^{-1}$ after LS2 (around 2027)~\cite{ohnishi2023ipac}. From a beam-beam perspective, the path to the target luminosity can be outlined by examining Eq.~(\ref{eq:xi2_by_lum}):
\begin{itemize}
    \item A total current of 1.4 A was achieved at LER in June 2022. A gain factor of 2.5 can be expected if the design value of 3.6 A is reached. Currently, the limit on total beam current is not from RF power but from the high risks of machine failures due to sudden beam losses~\cite{ikeda2022eefact}. Another obstacle to achieving higher beam currents in the future is that the LER beam's short lifetime requires strong injection power and good beam quality from the linac (see Ref.~\cite{funakoshi2023icfa} for details).
    \item A beam-beam parameter of 0.04 was achieved at LER. The beam-beam limit is expected to be around $\xi_y^L\sim 0.1$, suggesting a gain factor of 2.5. The important obstacles to achieving higher beam-beam parameters include 1) the vertical beam-size blowup driven by beam-beam and its interplay with nonlinear lattice~\cite{zhou2015icfa} and beam coupling impedances~\cite{lin2022prab, zhang2023prab}; 2) the vertical beam-size blowup in LER due to a so-called ``-1 mode instability'' that is driven by an interplay between the large vertical impedance (dominated by small-gap collimators) and the bunch-by-bunch feedback system (see Ref.~\cite{ishibashi2023icfa} for details); 3) the imperfect crab waist due to the non-transparent interaction region (IR)~\cite{morita2014ipac}. Installing a nonlinear collimator (it will replace a small-gap collimator and consequently reduce the vertical impedance significantly.) to the LER during LS1~\cite{natochii2022bg} is an expected solution for curing the vertical blowup due to collective effects. The effectiveness of crab waist relies on a clean IR, which means less nonlinearity in optics and less background to the Belle II. An R\&D program is ongoing for an IR upgrade after the LS2~\cite{itfir}.
    \item The vertical beta function reached $\beta_y^*=1$ mm. If the final design of $\beta_{y\pm}^*=0.27/0.3$ mm is achieved, it will result in a luminosity gain of 3.3. The important obstacles lie in the small dynamic aperture and short lifetime resulting from the nonlinear IR combined with beam-beam and crab waist~\cite{morita2014ipac}.
\end{itemize}

Assuming a balanced collision (i.e., $\beta_{y+}^*=\beta_{y}^*=\beta_{y}^*$ and $\epsilon_{y+}=\epsilon_{y-}=\epsilon_{y}$) and the hourglass effect is negligible, the incoherent beam-beam parameter Eq.~(\ref{eq:xi1_approx}) can be simplified as
\begin{equation}
    \xi_{y\pm}^i \approx
    \frac{r_e}{2\pi ef_0 \gamma_\pm \tan \frac{\theta_c}{2} }
    \frac{I_{b\mp}}{\sigma_{z\mp}}
    \sqrt{\frac{\beta_y^*}{\epsilon_y}}.
    \label{eq:xiy_symmetric_beam}
\end{equation}
Suppose a beam-beam limit of $\xi_y\sim 0.1$, we can examine the constraints on machine parameters:
\begin{itemize}
    \item With given total beam currents and $\beta_y^*$, to approach the beam-beam limit of $\xi_y\sim 0.1$, smaller $\epsilon_y$ is always preferred. With $\beta_y^*=1$ mm optics, achieving a single-beam emittance of $\epsilon_{y0} < 20$ pm at SuperKEKB has been challenging. However, the SuperKEKB design targets a vertical emittance of $\epsilon_y\sim 10$ pm with collective effects considered~\cite{Ohnishi2013PTEP, SuperKEKBTDR}.
    \item To keep $\xi_y \lesssim 0.1$, small $\beta_y^*$ is always preferred to allow higher beam currents. However, squeezing $\beta_y^*$ results in a smaller dynamic aperture and shorter lifetime.
\end{itemize}

\section{Summary}
\label{sec:summary}

Different notations have been used for the beam-beam parameters in evaluations of the luminosity performance in KEKB and SuperKEKB. The hourglass effects on the beam-beam parameters were counted or not in different cases. By the definition of Eq.~(\ref{eq:xi2_by_lum}), the beam-beam parameters achieved in SuperKEKB until June 2022 with crab waist at $\beta_y^*=1$ mm were 0.0565 and 0.0434 for LER and HER, respectively. These values are remarkably lower than the design values.

Luminosity tunings at SuperKEKB have been regularly done by following a well-defined routine from major optics corrections to minor knobs of IP parameters. The luminosity has been optimized by navigating a parameter space with many dimensions. 

Many factors contribute to the discrepancy in luminosity between simulation and measurement and remain to be investigated. The beam-beam interaction has been playing a key role in defining the luminosity performance at SueprKEKB. Machine operation experience suggests that reliable luminosity prediction via beam-beam simulations has been challenging at SuperKEKB. Multiple dynamics, such as beam-beam interaction, machine imperfections, impedance effects, etc., must be involved in simulations and call for the development of a general tracking code. Dedicated machine studies are also required to investigate the interplay of multiple dynamics at SuperKEKB.

\acknowledgments

We thank the SuperKEKB and the Belle II teams for their constant support of our work. Special thanks are due to Y. Zhang, K. Oide, D. Shatilov, M. Zobov, T. Browder, Y. Cai, K. Matsuoka, and S. Uehara for their fruitful discussions.

%\paragraph{Note added.} This is also a good position for notes added
%after the paper has been written.

% Bibliography
%% [A] Using JHEP.bst file
%\bibliographystyle{JHEP}
%\bibliography{biblio.bib}

%% or
%% [B] Manual formatting (see below)
%% (i) We suggest to always provide author, title and journal data or doi:
%% in short all the informations that clearly identify a document.
%% (ii) please avoid comments such as "For a review'', "For some examples",
%% "and references therein" or move them in the text. In general, please leave only references in the bibliography and move all
%% accessory text in footnotes.
%% (iii) Also, please have only one work for each \bibitem.

\end{document}